\documentclass{elsart}

\usepackage{graphicx}
\usepackage{amssymb,amsmath}
\usepackage{ae}

\begin{document}

\begin{frontmatter}



\title{Statistical analysis of the distribution of amino acids in {\em Borrelia burgdorferi} genome under different genetic codes}


\author[ulsa,iib]{Jos\'e A. Garc\'ia\corauthref{cor}},
\corauth[cor]{Corresponding address: Laboratory of Theoretical Biology, Research Department, La Salle University, Benjamin Franklin 47, Col. Hipodromo-Condesa, Mexico D.F. 06140, Mexico, Fax: (52-55)5515-7631}
\ead{jgarcia@ci.ulsa.mx}
\author[ulsa]{Samantha Alvarez},
\author[ulsa]{Alejandro Flores},
\author[iib]{Tzipe Govezensky},
\author[iib]{Juan R. Bobadilla},
\author[iib]{Marco V. Jos\'e}

\address[ulsa]{Research Department, La Salle University,  Benjamin Franklin 47,\\
     Col. Hipodromo-Condesa, M\'exico, D.F. 06140, M\'exico.}

\address[iib]{Instituto de Investigaciones Biom\'edicas, Universidad  Nacional\\
  Aut\'onoma de M\'exico, A.P. 70228, M\'exico D.F. 04510, M\'exico}

\thanks{MVJ was financially supported by PAPIIT IN205702, UNAM, M\'exico.}

\begin{abstract}
% Text of abstract
The genetic code is considered to be universal. In order to test if some statistical properties of the coding bacterial genome were due to inherent properties of 
the genetic code, we compared the autocorrelation function, the scaling properties and the maximum entropy of the distribution of distances of amino acids in sequences obtained by translating protein-coding regions from the genome of {\em Borrelia burgdorferi}, under different genetic codes. Overall our results indicate that these properties are very stable to perturbations made by altering the genetic code. We also discuss the evolutionary likely implications of the present results.

\end{abstract}

\begin{keyword}
% maximum of four
Genomics \sep genetic codes \sep statistical analysis

\PACS  87.10.+e \sep 05.40.+j
% See http://www.aip.org/pacs/pacs.html
% or  http://www.elsevier.com/locate/pacs

\end{keyword}

\end{frontmatter}

\section{Introduction}
Organisms use the genetic code to translate the information stored in DNA or RNA nucleotide sequences to synthesize amino acids sequences called proteins.  The same code is used in all living organisms (there is an exception in the mitochondrial genome), so it is nearly universal \cite{lewin}.

The universality of the genetic code suggests that it should have been established early in evolution, so once it appeared in nature it was ``frozen'' \cite{crick68}. An unanswered question is whether other genetic codes could accomplish the same function or be as efficient as the actual universal genetic code.

In order to test the properties of different genetic codes, we analyzed some discriminating statistics of the distance distribution of amino acids (aa) derived from protein-coding regions from the genome of {\em Borrelia burgdorferi}, through numerical experiments in which the  actual genetic code was perturbed. The chosen statistics are related to the content of information, the scaling properties of the distances series, and the autocorrelation properties.

\section{Experimental design}
\label{design}
We started with a sequence of protein-coding regions of the genome of {\em Borrelia burgdorferi} (see \cite{bis}). This sequence was translated to a sequence of aa using different genetic codes (see below). For the three stop codons, we assigned character $X$. Then, we generated distance series between identical aa along the chromosome for each character (either aa or an stop codon).  Our master control was the universal genetic code itself. As negative controls, we considered both a shuffled version of the original sequence (shuffled code), and a synthetic DNA sequence $10^{6}$ nucleotides long, obtained by  sampling the four DNA nucleotides with replacement (random code).

Instead of the classic random walk mapping of a DNA sequence \cite{li92,peng92}, we followed a different approach for studying the statistical properties of aa sequences. In particular, for a given aa, we determined its actual position along the whole sequence, and from this we measured, as distance, the number of aa which lies between two identical characters. Hence, we obtained the actual distance series for each character.

\subsection{Genetic codes}

The universal code presents a characteristic distribution of codons to aa. In this distribution, there are several aa which are encoded by more than one codon, so it is degenerated \cite{lewin}. Often the base in the third position is less significant,  as a mutation in this position does not imply a change in the encoded aa (third-base degeneracy).
 
The first perturbation was called code 2. Amino acids were randomly assigned  to codons, preserving the universal distribution of codons to aa i.e. the degeneracy of the code is the same as the universal code. In this way, a mutation in the third base altered the coded aa, thus the third-base degeneracy is not longer sustained.

In the uniform code we assumed a uniform distribution of codons to aa, so each aa is coded by three different randomly chosen codons. As there are 20 aa, the uniform code has four stop codons. 

For generating what we called the crazy code, we built a population, in which the 21 characters (representing either an aa or an stop codon) were sampled with replacement, and each of them was randomly assigned to one out of the 64 codons, so the distribution of codons to aa is both not universal-like, and not uniform, e.g. three different aa can be translated with five different codons.

Finally, we also tested a perturbation following the RNA world hypothesis \cite{gilbert86} using the RNY (purine-any nucleotide-pyrimidine) pattern proposed by Eigen \& Schuster \cite{eigen79}. In the RNA world code aspartic acid is coded by GAC and GAU; asparagine is coded by AAC and AAU; alanine is coded by GCU and GCC; isoleucine is coded by AUU and AUC; glycine is coded by GGC and GGU; serine is coded by AGC and AGU; threonine is coded by ACU and ACC; and, valine is coded by GUU and GUC. This code has been proposed as the primeval genetic code (see also \cite{konecny95}).

\subsection{Statistical analysis}
In order to test for statistical differences among the master, the negative and the synthetic codes (code 2, uniform code, crazy  code and RNA world code), both the $p$-value of the Wilcoxon-Mann-Withney test \cite{mann47} and the bootstrap \cite{efron98} 95\% confidence intervals (C.I. 95)  were calculated for three different statistics. The first technique is used to test differences of means regardless of the particular distribution of the random variable and the latter estimates non-parametric confidence intervals by sampling few data several times (e.g. 1000 times) with replacement. The bootstrap C.I. 95 from the random code gives an estimation of the white noise bandwidths.

The chosen statistics were: a) The average of the mean of the first 38 lags of the autocorrelation function (ACF); b) Mean of the detrended fluctuation analysis scaling exponent (DFA)\cite{dfa}; c) Average of the geometric variation coefficient maximum entropy ($ME-gvc$), where $gvc$ is the ratio of 
 $sd(x)/\tilde{x}$, where $sd(x)$ is the standard deviation of the maximum entropy of the series $x$, and $\tilde{x}$ is the geometric mean of the maximum entropy of the series $x$. Maximum entropy of the series was calculated with the {\em SSA-MTM Toolkit}, v4.2 \cite{mem}. All means were calculated for the 21 characters within each code.
 
Both the ACF and the DFA look for autocorrelations within the series. The DFA technique is based on a modified root mean square analysis of a random walk, to assess the intrinsic correlation properties of a dynamic system separated from external trends in the data, and is intended to determine the scaling properties of a time series \cite{dfa}. When the DFA calculated scaling exponent is equal to 0.5 it is indicative of white noise; if the value lies betweeen 0.5 and 1 then the time series exhibits long-range correlations.

\section{Results}
\label{results}
\begin{figure}[ht]
\begin{center}
\includegraphics{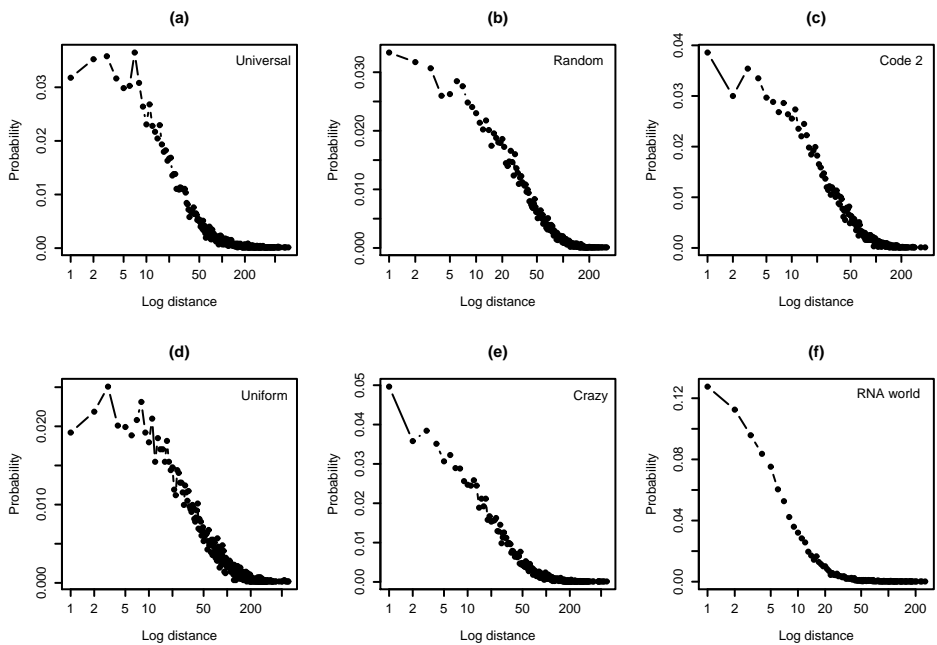}
\caption{\label{dist}Probability density functions of the distance distribution of aspartic acid. Distance is measured as the number of aa that are between two identical aa. a) Universal code; b) Random code; c) Code 2; d) Uniform code; e) Crazy code; f) RNA world code.}
\end{center}
\end{figure}

The distance distributions for each particular aa changed for all the tested codes. Fig. \ref{dist} shows the probability density function (pdf) of the distance distribution for aspartic acid as an example. Interestingly, the pdfs of various aa obtained with different codes presented an oscillatory decaying pattern. The only exception to this pattern was observed in the RNA world code, in which none of the aa presented oscillations.
 
Several works have reported periodical patterns in DNA sequences by means of autocorrelation function (ACF) analysis  \cite{shepherd81,arques90,karlin93,arques97,herzel97}. 
We also utilized the ACF, but applied to the aa distance series, coming from both the universal code and the negative controls (shuffled and random codes). We found autocorrelations in the universal code, and  no autocorrelation for all lags in the negative controls (as indicated by the bandwidth of white noise). As can be seen from Table \ref{acftab}, there is a clear difference, as measured by the bootstrap C.I. 95
 between the universal code and the negative controls. Further, we tested if the average of the mean of the first 38 lags of the ACF of the universal code, was statistically different to any of the other codes. We found differences with code 2, crazy code, and uniform code with a non-parametric test. These synthetic codes distributions were displaced to lower values to respect of the universal code distribution, although with an slight overlapping (see Table \ref{acftab}).

\begin{table}[htdp]
\caption{Bootstrap confidence intervals (C.I.) for: autocorreletion function analysis (ACF); detrended fluctuation analysis scaling exponent (DFA); and geometric variation coefficient of maximum entropy (ME-gvc).}
\begin{tabular}{lccc}
\hline
Code & C.I. (ACF) & C.I. (DFA) & C.I. (ME-gcv)\\
\hline
Universal & 0.09 -- 0.12 & 0.72 -- 0.75 & 0.87 -- 1.14\\
Shuffled & 0.03 -- 0.06$^{\ddagger}$ & 0.56 -- 0.59$^{\ddagger}$ & 0.23 -- 0.46$^{\ddagger}$\\
Random & -1.5e-3 -- 1.2e-7$^{\ddagger}$ & 0.49 -- 0.51$^{\ddagger}$ & 0.07 -- 0.09$^{\ddagger}$\\
Code 2 & 0.07 -- 0.09$^{\dagger}$ & 0.69 -- 0.73$^{*}$ & 0.61 -- 0.84$^{\dagger}$\\
Crazy & 0.06 -- 0.10$^{*}$ & 0.69 -- 0.73$^{*}$ & 0.62 -- 0.90$^{\dagger}$\\
Uniform & 0.07 -- 0.10$^{*}$ & 0.71 -- 0.74 & 0.67 -- 0.89$^{*}$\\
RNA World & 0.06 -- 0.13 & 0.65 -- 0.73$^{*}$ & 0.51 -- 1.11\\
\hline
\multicolumn{3}{l}{$^{*}$p < 0.05, $^{\dagger}$p < 0.01, $^{\ddagger}$p < 0.001 (Wilcoxon-Mann-Withney test)}\\
\end{tabular}
\label{acftab}

\end{table}

Several authors have reported long-range correlations in DNA
\cite{peng92,voss92,peng94,arneodo95,mohanty00}. Here, in order to look for long-range correlations between each aa, we calculated the DFA scaling exponent \cite{dfa} for each distance series, and then tested for differences in the mean value for each code against the corresponding value obtained with the universal code. We found statistical differences with code 2, crazy code and the RNA world code, although with slight overlaps  in the bootstrap C.I. 95 in all cases (see Table \ref{acftab}). As expected, the DFA scaling exponents of the negative controls lie within (random) or very close (shuffled) to values that indicate brownian motion. It is worth to mention, that both negative controls are strikingly different when compared with all other codes.

Entropy, as a measure of information, has also been used to analyze DNA and to make comparisons  between coding and non-coding regions \cite{mantegna94,havlin95,schmitt97}. In the current study, we calculated the average of ME-gvc for each distance series. Again there was a clear difference between the statistics of the negative controls, and all the other codes. Statistical differences, were also found with code 2, crazy code and uniform code against the universal code with slight overlap in the bootstrap C.I. 95 (see Table \ref{acftab}).

\section{Discusion}
\label{discuss}

There have been several papers, some of them considered as classics, which have contributed to our understanding of the origin of the genetic code \cite{gilbert86,eigen79,crick68,konecny93,beland94,hartman94,konecny95}. However, none of them addressed the statistical properties of the translated sequence of aa. Here we carried out numerical experiments with different genetic codes in which some statistical properties of the translated products are analyzed. In order to study the coding DNA, we based our analysis on sequences of aa obtained by translating the protein coding sequence from {\em Borrelia burgdorferi} genome.

Peng, {\em et al.} \cite{peng92} have found that noncoding DNA sequences show long-range autocorrelations whereas coding sequences do not. This is remarkable in the case of bacterial chromosomes since most of the DNA content is coding. Indeed, they showed that in bacteria there is a lack of autocorrelation. We found autocorrelation in both DNA coding sequences \cite{bis} and in aa sequences coming from translating bacterial coding DNA. These apparently  contradictory results are presumably due to differences in the experimental design, as we looked for the distance series between characters (either aa or $n-$tuples of DNA).

In general the bootstrap C.I. 95 for all the tested statistics of the synthetic codes, showed a diminution of information content, weaker long-range correlations, and smaller values of the scaling exponent, when compared with the master code. Then, the universal code seems to contain optimum values for those statistics.

Regardless of finding statistical differences with alternative codes, in all cases the statistics have values closer to the universal code than to the negative controls. This suggests that the genetic code is very robust to perturbations, as information measures, as well as long and short correlations are maintained. Thus, once the universal code was established, it became fixed and resistant to evolutionary changes. The question of what makes unique the universal code remains an unanswered problem.

% \appendix

% \section{}
% \label{}

% Bibliographic references with the natbib package:
% Parenthetical: \citep{Bai92} produces (Bailyn 1992).
% Textual: \citet{Bai95} produces Bailyn et al. (1995).
% An affix and part of a reference:
%   \citep[e.g.][Ch. 2]{Bar76}
%   produces (e.g. Barnes et al. 1976, Ch. 2).


\begin{thebibliography}{99}

% \bibitem[Names(Year)]{label} or \bibitem[Names(Year)Long names]{label}.
% (\harvarditem{Name}{Year}{label} is also supported.)

\bibitem{lewin}{B. Lewin, Genes VII (2000) New York, Oxford University Press.}

\bibitem{crick68}{F.H.C. Crick, J. Mol. Biol. 38 (1968) 367. }

\bibitem{bis}{J. S\'anchez, M.V. Jos\'e, Biochem. Biophys. Res. Comm. 299 (2002) 126.}

\bibitem{li92}{W. Li, K. Kaneko, Europhys. Lett. 17 (1992) 655.}

\bibitem{peng92}{C.-K. Peng, S.V. Buldyrev, A.L. Goldberger, S. Havlin, F. Sciortino, M. Simons, H.E. Stanley, Nature 356 (1992) 168.}

\bibitem{gilbert86}{W. Gilbert, Nature 319 (1986) 618.}

\bibitem{eigen79}{M. Eigen, P. Schuster, The hypercycle: A principle of self-organization (1979) Berlin, Springer-Verlag.}

\bibitem{konecny95}{J. Konecny, M. Sch\"oniger, G.L. Hofacker, J. theor. Biol. 173 (1995) 263.}

\bibitem{mann47}{H.B. Mann, D.R. Whitney, Ann. Math. Statist. 18 (1947) 50.}

\bibitem{efron98}{B. Efron, R.J. Tibshirani, An introduction to the bootstrap (1998) San Francisco, Chapman \& Hall.}

\bibitem{dfa}{C-K. Peng, S. Havlin, H.E. Stanley, A.L. Goldberger, Chaos 5 (1995) 82.}

\bibitem{mem}{M. Ghil, M.R. Allen, M.D. Dettinger, K. Ide, D. Kondrashow, M.E. Mann, A.W. Robertson, A. Saunders, Y. Tian, F. Varadi, P. Yiou, Rev. Geophys. 40 (2002) 1.}

\bibitem{shepherd81}{J.C.W. Shepherd, J. Mol. Evol. 17 (1981) 94.}

\bibitem{arques90}{D. G. Arqu\`es, C.J. Michel, J. theor. Biol. 143 (1990) 307.}

\bibitem{karlin93}{S. Karlin, V. Brendel, Science 259 (1993) 677.}

\bibitem{arques97}{D.G. Arqu\`es, C.J. Michel, BioSystems 44 (1997) 107.}

\bibitem{herzel97}{H. Herzel, I.Gro$\beta$e, Phys. Rev. E 55 (1997) 800.}

\bibitem{voss92}{R.F. Voss, Phys. Rev. Lett. 68 (1992) 3805.}

\bibitem{peng94}{C.-K. Peng, S.V. Buldyrev, S. Havlin, M. Simmons, H.E. Stanley, A.L. Goldberger, Phys. Rev. E 49 (1994) 1685.}

\bibitem{arneodo95}{A. Arneodo, E. Bacry, P.V. Graves, J.F. Muzy, Phys. Rev. Lett. 74 (1995) 3293.}

\bibitem{mohanty00}{A.K. Mohanty, A.V.S.S.N. Rao, Phys. Rev. Lett. 84 (2000) 1832.}

\bibitem{mantegna94}{R.N. Mantegna, S.V. Buldyrev, A.L. Goldberger, S. Havlin, C.-K. Peng, M. Simons, H.E. Stanley, Phys. Rev. Lett. 73 (1994) 3169.}

\bibitem{havlin95}{S. Havlin, S.V. Buldyrev, A.L. Goldberger, R.N. Mantegna, C.-K. Peng, M. Simons, H.E. Stanley, Fractals 3 (1995) 269.}

\bibitem{schmitt97}{A.O. Schmitt, H. Herzel, J. theor. Biol. 188 (1997) 369.}

\bibitem{konecny93}{J. Konecny, M. Eckert, M. Sch\"oniger, G.L. Hofacker, J. Mol. Evol. 36 (1993) 407.}

\bibitem{beland94}{P. B\'eland, T.F.H. Allen, J. theor. Biol. 170 (1994) 359.}

\bibitem{hartman94}{H. Hartman, J. Mol. Evol. 40 (1995) 541.}

\end{thebibliography}
\end{document}